\documentclass{EslabStyle} % remove later
\usepackage{epsfig}

\begin{document}

\def\arcs{\ifmmode {''}\else $''$\fi}
\def\ion#1#2{#1$\;${\sc{#2}}\relax}

\def\aj{\rm{AJ}}                   % Astronomical Journal
\def\araa{\rm{ARA\&A}}             % Annual Review of Astron and Astrophys
\def\apj{\rm {ApJ}}                % Astrophysical Journal
\def\apjl{\rm{ApJ}}                % Astrophysical Journal, Letters
\def\apjs{\rm{ApJS}}               % Astrophysical Journal, Supplement
\def\apss{\rm{Ap\&SS}}             % Astrophysics and Space Science
\def\aap{\rm{A\&A}}                % Astronomy and Astrophysics
\def\aapr{\rm{A\&A~Rev.}}          % Astronomy and Astrophysics Reviews
\def\aaps{\rm{A\&AS}}              % Astronomy and Astrophysics, Supplement
\def\mnras{\rm{MNRAS}}             % Monthly Notices of the RAS
\def\nat{\rm{Nature}}              % Nature
\def\procspie{\rm{Proc.~SPIE}}     % Proceedings of the SPIE

\title{Infrared Spectroscopy of a Massive Obscured Star Cluster in the
Antennae Galaxies (NGC 4038/4039) with NIRSPEC}

\author{A.\,M. Gilbert\inst{1} \and J.\,R. Graham\inst{1} \and I.\,S.
  McLean\inst{2} \and E.\,E. Becklin\inst{2} \and D.\,F. Figer\inst{3}
  \and J.\,E. Larkin\inst{2} \and N.\,A. Levenson\inst{4} \and H.\,I.
  Teplitz\inst{5,6} \and M.\,K. Wilcox\inst{2}} \institute{Department
  of Astronomy, University of California -- Berkeley, CA 94720-3411,
  USA \and Department of Physics and Astronomy, University of
  California -- Los Angeles, CA, 90095-1562, USA \and Space Telescope
  Science Institute -- 3700 San Martin Dr., Baltimore, MD 21218, USA
  \and Department of Physics and Astronomy, Johns Hopkins University
  -- Baltimore, MD 21218, USA \and Laboratory for Astronomy and Solar
  Physics -- Code 681, Goddard Space Flight Center, Greenbelt MD
  20771, USA \and NOAO Research Associate }

\maketitle 

\begin{abstract}
We present infrared spectroscopy of the Antennae Gal\-axies
(NGC~4038/4039) with NIRSPEC at the W. M. Keck Observatory.  We imaged
the star clusters in the vicinity of the southern nucleus (NGC~4039)
in $0\arcs.39$ seeing in K-band using NIRSPEC's slit-viewing
camera. The brightest star cluster revealed in the near-IR (M$_{\rm
K}(0) \simeq -17.9$) is insignificant optically, but coincident with
the highest surface brightness peak in the mid-IR ($12-18 \mu$m) ISO
image presented by \cite*{mirabel98}.  We obtained high
signal-to-noise 2.03$-$2.45 $\mu$m spectra of the nucleus and the
obscured star cluster at R $\sim 1900$.

The cluster is very young (age $\sim 4$ Myr), massive (M $\sim 16
\times 10^6$ M$_{\odot}$), and compact (density $\sim 115$ M$_{\odot}$
pc$^{-3}$ within a 32 pc half-light radius), assuming a Salpeter IMF
(0.1$-$100 M$_{\odot}$).  Its hot stars have a radiation field
characterized by T$_{\rm eff}\sim 39,000$ K, and they ionize a compact
\ion{H}{ii} region with n$_{\rm e}\sim 10^4$ cm$^{-3}$.  The stars are
deeply embedded in gas and dust (A$_{\rm V} \sim 9-10$ mag), and their
strong FUV field powers a clumpy photodissociation region with
densities n$_{\rm H}\ga 10^5$ cm$^{-3}$ on scales of $\sim 200$ pc,
radiating L$_{{\rm H}_2 1-0~{\rm S}(1)}= 9600$ L$_{\odot}$.

\end{abstract}

\section{Introduction}

The Antennae (NGC~4038/4039) are a pair of disk galaxies in an early
stage of merging which contain numerous massive super star clusters
(SSCs) along their spiral arms and around their interaction region
(\cite{whitmore95,whitmore99}).  The molecular gas distribution peaks
at both nuclei and in the overlap region (\cite{stanford90}), but the
gas is not yet undergoing a global starburst typical of more advanced
mergers (\cite{nikola98}).  Star formation in starbursts appears to
occur preferentially in SSCs.  We chose to observe the Antennae
because their proximity permits an unusually detailed view of the
first generation of merger-induced SSCs and their influence on the 
surrounding interstellar medium.

The Infrared Space Observatory (ISO) $12-18~\mu$m image showed that
the hot dust distribution is similar to that of the gas, but peaks at
an otherwise inconspicuous point on the southern edge of the overlap
region (\cite{mirabel98}; see \cite{wilson00}).  This powerful
starburst knot is also a flat-spectrum radio continuum source
(\cite{hummel86}) and may be associated with an X-ray source
(\cite{fabbiano97}).  We imaged the region around this knot, and
discovered a bright compact star cluster coincident with the mid-IR
peak.  We obtained moderate-resolution (R $\sim 1900$) K-band spectra
of both the obscured cluster and the NGC~4039 nucleus.

\begin{figure}[t]
  \begin{center}
    \epsfig{file=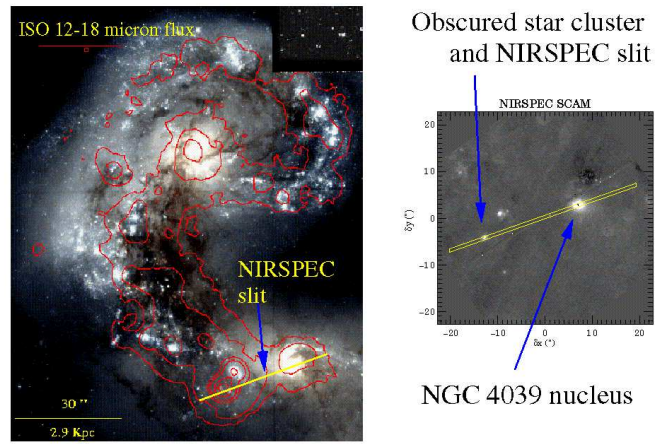,width=3.6in}
  \end{center}
%\resizebox{\hsize}{!}{\includegraphics{gravityRGBs.ps}}
\caption{HST image of the Antennae overlaid with ISO 12$-$18 $\mu$m
contours (\protect\cite{mirabel98}), and NIRSPEC K band image showing the slit
position and several star clusters in the field.}
\label{scam}
\end{figure}

\section{Observations \& Data Reduction}
NIRSPEC is a new facility infrared ($0.95 - 5.6 \mu$m) spectrometer
for the Keck-II telescope, commissioned during April through July,
1999 (\cite{mclean98}).  It has a cross-dispersed cryogenic echelle
with R $\sim 25,000$, and a low resolution mode with R $\sim 2000$.
The spectrometer detector is a 1024 $\times$ 1024 InSb ALADDIN focal
plane array, and the IR slit-viewing camera detector is a 256 $\times$
256 HgCdTe PICNIC array.

\begin{figure}[t]
  \begin{center}
    \epsfig{file=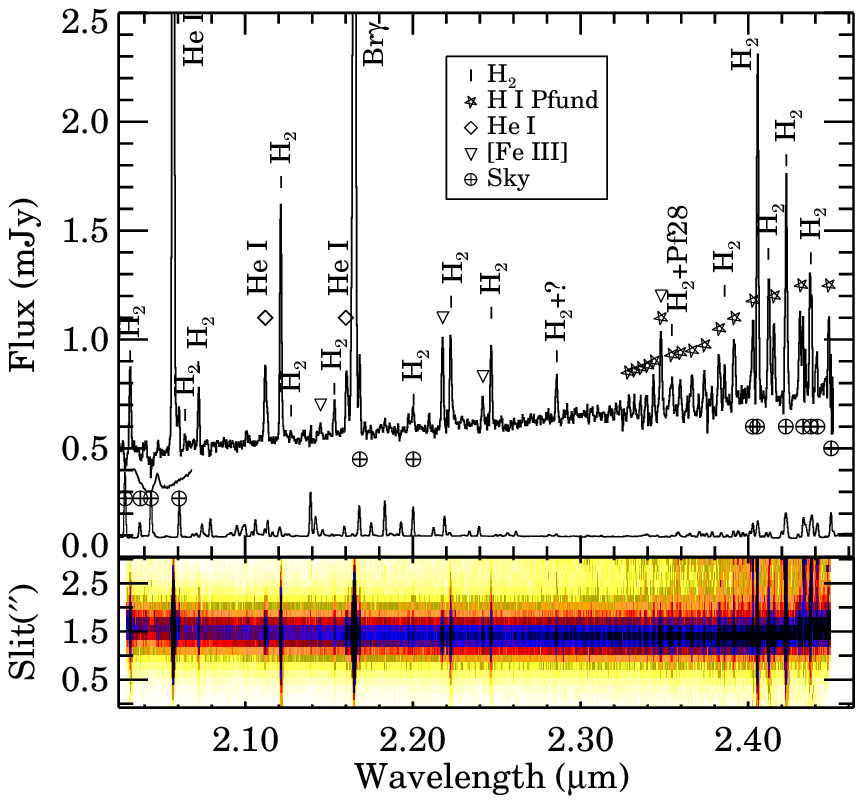,width=3.35in}
  \end{center}
%\resizebox{\hsize}{!}{\includegraphics{gravityRGBs.ps}}
\caption{NIRSPEC spectrum of the obscured star
cluster shows nebular and fluorescent H$_2$ emission with a continuum
rising toward the red.  Scaled sky counts are plotted at 0.1 mJy.
$\omega$-shaped curve represents an atmospheric CO$_2$ band at 2.05
$\mu$m.
\label{cluster}}
\end{figure}

\begin{figure}[t]
  \begin{center}
    \epsfig{file=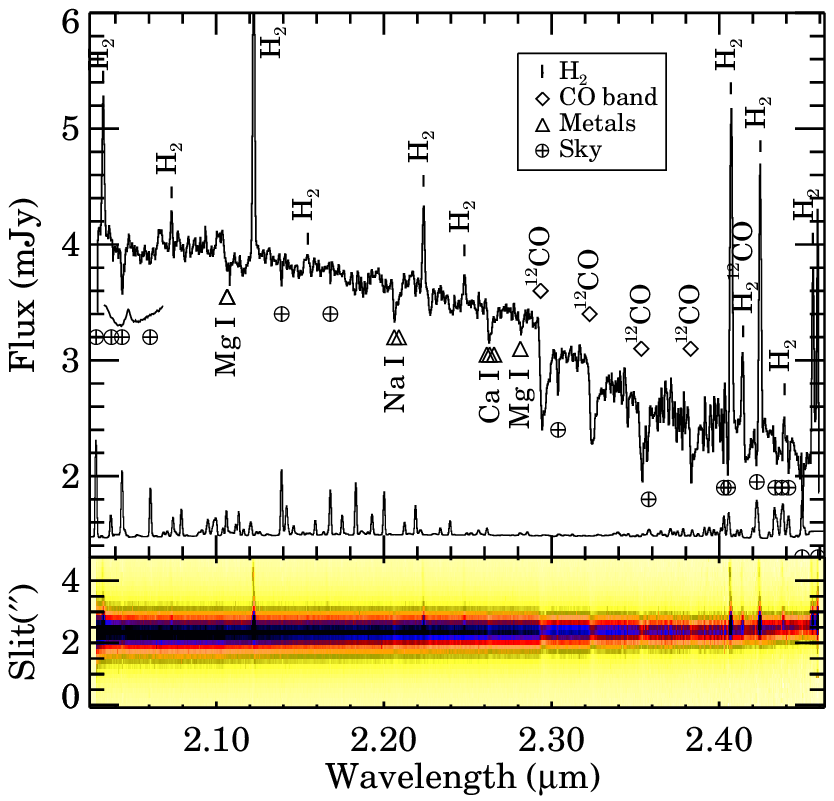,width=3.35in}
  \end{center}
%\resizebox{\hsize}{!}{\includegraphics{gravityRGBs.ps}}
\caption{NIRSPEC spectrum of NGC~4039 nucleus shows
extended collisionally excited H$_2$ emission and a strong stellar
continuum marked by photospheric absorption.  No Br$\gamma$ is present.
Scaled sky counts are shown at 1.5 mJy.
\label{nucleus}}
\end{figure}

We observed the Antennae with NIRSPEC during the June 1999
commissioning run.  Figure~\ref{scam} shows the HST and ISO image of
\cite*{mirabel98} together with NIRSPEC slit-viewing camera (SCAM)
images at 2 $\mu$m; the SCAM images reveal that the mid-IR ISO peak is
a bright (K $= 14.6$) compact star cluster located 20\arcs.4 east and
4\arcs.7 north of the K-band nucleus.  This cluster is associated with
a faint (V $= 23.5$) red (V$-$I $= 2.9$) source (\# 80 in
\cite{whitmore95}) visible with Space Telescope (Whitmore \& Zhang,
private communication).  We obtained low resolution (R $\simeq 1900$)
$\lambda \simeq 2.03 - 2.45~\mu$m spectra through a $0\arcs.57 \times
42\arcs$ slit at PA=77$^{\rm o}$ located on the obscured star cluster
and the nucleus of NGC~4039, for a total integration time of 2100 s.
We reduced the data following the standard procedure, described in
more detail by \cite*{gilbert00}.  Reduced spectra are shown in
Figures~\ref{cluster} and \ref{nucleus}.

\section{Massive Star Cluster}
\label{sec:cluster}

The cluster spectrum is characterized by strong emission
lines\footnote{A table of measured line fluxes is available
electronically from 
%url
{http://astro.berkeley.edu/$\sim$agilbert/antennae}.} and
a continuum (detected with SNR $\simeq 15$) dominated by the light of
hot, blue stars and dust.  The nebular emission lines are slightly
more extended than the continuum, and the H$_2$ emission is even more
extended.  This suggests a picture in which hot stars and dust are
embedded in a giant compact H~{\sc ii} region surrounded by clumpy
(see \S 3.2) clouds of obscuring gas and dust whose surfaces are
ionized and photodissociated by FUV photons escaping from the star
cluster.

For a distance to the Antennae of 19 Mpc (H$_0$=75 km s$^{-1}$
Mpc$^{-1}$, 1\arcs = 93 pc) (\cite{whitmore99}), we find that the
cluster has M$_{\rm K} = -16.8$.  We estimate the screen extinction to
the cluster by assuming a range of (V$-$K)$_0 \simeq 0-1$ as expected
from Starburst99 models (\cite{leitherer99}), and that A$_{\rm K} =
0.11$ A$_{\rm V}$ (\cite{rieke85}).  We find A$_{\rm V} = 9-10$ mag,
which implies M$_{\rm K}(0) = -17.9$, adopting A$_{\rm K}=1.1$ (which
is confirmed by our analysis of the \ion{H}{ii} recombination lines in
\S 3.1).  We can use the intrinsic brightness along with the Lyman
continuum flux inferred from the de-reddened Br$\gamma$ flux ($3.1
\times 10^{-14}$ erg s$^{-1}$ cm$^{-2}$), Q(H$^+)_0 = 1.0 \times
10^{53}$ photons s$^{-1}$, to constrain the cluster mass and age.
Using instantaneous Starburst99 models we find a total mass of $\sim 7
\times 10^6$ M$_{\odot}$ (with $\sim 2600$ O stars) for a Salpeter IMF
extending from 1 to 100 M$_\odot$, and an age of $\sim 4$ Myr.  This
age is consistent with the lack of photospheric CO and metal
absorption lines from red supergiants and other cool giants, which
would begin to contribute significantly to the 2 $\mu$m light at an
age of $\sim 7$ Myr (\cite{leitherer99}).  The cluster's density is
then about 115 M$_{\odot}$ pc$^{-3}$ for stars of 0.1$-$100
M$_{\odot}$ within a half-light radius of $32$ pc.  This density is 30
times less than that of the LMC SSC, R136 (within a radius of 1.7 pc,
assuming a Salpeter proportion of low-mass stars) (\cite{hunter95}).
Thus the Antennae cluster may be a complex of clusters rather than one
massive cluster.

\subsection{Nebular Emission}

The cluster spectrum features a variety of nebular lines that reveal
information about the conditions in and around the gas ionized gas by
the cluster, which in turn allows us to constrain the effective
temperature of the ionizing stars.

\begin{figure}[t]
  \begin{center}
    \epsfig{file=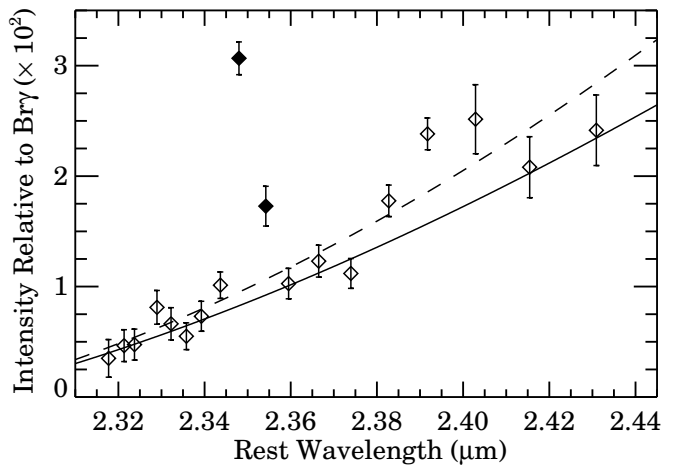}
  \end{center}
%\resizebox{\hsize}{!}{\includegraphics{gravityRGBs.ps}}
\caption{Pfund line fluxes relative to Br$\gamma$
flux ($1.05 \times 10^{-14}$ erg s$^{-1}$cm$^{-2}$).  Solid curve is
unextincted theoretical curve for n$_{\rm e} = 10^{4}$ cm$^{-3}$,
T$_{\rm e}=10^{4}$ K (Hummer \& Storey 1987).  Filled symbols
represent lines that are known blends, and the dashed curve shows
theoretical fluxes with the best-fit extinction A$_{\rm K}=1.1$
mag.
\label{pfund}} 
\end{figure}

We detected \ion{H}{i} Pfund series lines from Pf~19 to Pf~38, which
allow us to infer the extinction across the K window.  We display the
Pfund fluxes relative to that of Br$\gamma$ in Figure~\ref{pfund},
where filled symbols give fluxes for the blends Pf 28+H$_2$ 2$-$1 S(0)
and Pf 29+[\ion{Fe}{iii}].  They fall well above the other points,
which follow closely the theoretical expectation for intensities
relative to Br$\gamma$ (solid curve) with no reddening applied, for a
gas with n$_{\rm e} = 10^{4}$ cm$^{-3}$ and T$_{\rm e}=10^{4}$ K
(\cite{hummer87}).  Excluding the two known blends, the best-fit
foreground screen extinction is A$_{\rm K}=1.1 \pm 0.3 $ mag (dashed
curve), assuming the extinction law of \cite*{landini84} and evaluated
at 2.2 $\mu$m.  We consider this an upper limit on A$_{\rm K}$ because
a close look at the spectrum shows that the points above the dashed
line in Figure~\ref{pfund} for Pf 22$-$24 at 2.404, 2.393, and 2.383
$\mu$m may also be blended or contaminated by sky emission, implying a
lower A$_{\rm K}$ and a much better fit to the theory.  Hence the
majority of the extinction to the cluster is bypassed by observing it
in K band.

The ratios of [\ion{Fe}{iii}] 2.146 $\mu$m and 2.243 $\mu$m lines to
[\ion{Fe}{iii}] 2.218 $\mu$m are nebular density diagnostics,
consistent with a fairly high density, n$_{\rm e}= 10^{3.5} - 10^4$
cm$^{-3}$.  (See \cite*{gilbert00} for more detail.)
The \ion{He}{i} line ratios can be used to infer nebular temperature
T$_{\rm e}$, and are fairly insensitive to n$_{\rm e}$.  We find the
ratio \ion{He}{i} 2.1128+2.1137 $\mu$m/\ion{He}{i} 2.0589 $\mu$m = 0.052
$\pm$ 0.003, which for n$_{\rm e}$=10$^4$ cm$^{-3}$ is consistent with
T$_{\rm e}=17,500 \pm 800$ K (\cite{benjamin99}).  However this is much
hotter than typical nebular temperatures, and may indicate some
non-nebular contribution from hot stars to the line emission.

The \ion{He}{i} 2.0589 $\mu$m/Br~$\gamma$ ratio is an indicator of the
T$_{\rm eff}$ of hot stars in \ion{H}{ii} regions (\cite{doyon92}),
although it is sensitive to nebular conditions such as the relative
volumes and ionization fractions of He$^+$ and H$^+$, geometry,
density, dustiness, etc. (\cite{shields93}). \cite*{doherty95} studied
H and He excitation in a sample of starburst galaxies and \ion{H}{ii}
regions.  For starbursts they found evidence for high-T$_{\rm eff}$,
low-n$_{\rm e}$ ($\sim 10^2$ cm$^{-2}$) ionized gas from \ion{He}{i}
2.0589 $\mu$m/Br~$\gamma$ ratios of 0.22 to 0.64.  The ultra-compact
\ion{H}{ii} regions were characterized by higher ratios (0.8$-$0.9)
and higher densities, $\sim$ 10$^4$ cm$^{-3}$.  The cluster has a flux
ratio of 0.70, a value between the two object classes of
\cite*{doherty95}.  Assuming the line emission is purely nebular, this
ratio is consistent with a high-density (10$^4$ cm$^{-3}$) model of
\cite*{shields93} (also indicated by the [\ion{Fe}{iii}] emission), and
implies T$_{\rm eff}\simeq 39,000$ K for the assumed model parameters.

The cluster has properties more like those of a compact \ion{H}{ii}
region than a diffuse one.  It appears to be a young, hot,
high-density \ion{H}{ii} region, one of the first to form in this part
of the Antennae interaction region.

\subsection{Molecular Emission}

The spectrum shows evidence for almost pure UV fluorescence excited by
FUV radiation from the O \& B stars; the strong, vibrationally excited
1$-$0, 2$-$1 \& 3$-$2 H$_2$ emission has T$_{\rm vib} \ga 6000$~K and
T$_{\rm rot}\simeq 970$, 1600, and 1800 K, respectively, and weak
higher-v (6$-$4, 8$-$6, 9$-$7) transitions are present as well.  The
H$_2$ lines are extended over $\simeq 200$~pc, about twice the extent
of the continuum and nebular line emission, so a significant fraction
of the FUV (912$-$1108 \AA) light escapes from the cluster to heat and
photodissociate the local molecular ISM.

\begin{figure}[th]
\begin{center}
\epsfig{file=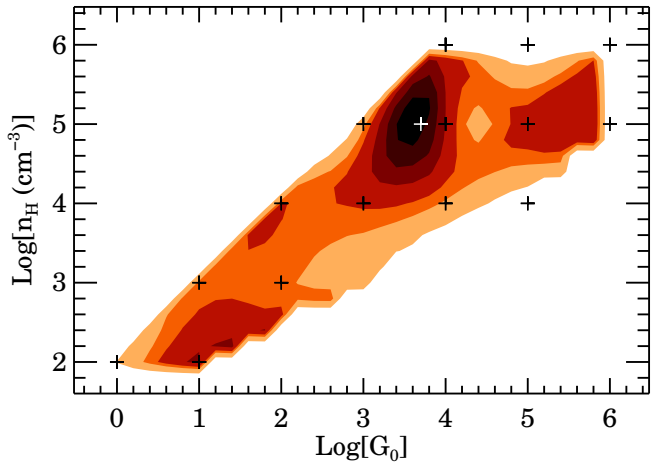}
\end{center}
\caption{Comparison of H$_2$ line strengths with PDR models.  Contours
of $\chi^2_\nu$ for 22 lines projected onto n$_{\rm H} -$G$_0$ plane
peak at n$_{\rm H} \sim 10^5$ cm$^{-3}$ and G$_0 \sim 5000$. Model
points (+) are for T$_0$ = 300 $-$ 2000 K. White + marks best-fit PDR
model of Draine \& Bertoldi (1996), with T$_0$ = 900 K and
$\chi_\nu^2=9.3$.  Contours are 50, 25, 20, 15, 12, 10.
\label{chisq}
}
\end{figure}

We compared the photodissociation region (PDR) models of
\cite*{draine96} with our data by calculating reduced $\chi_\nu^2$.
Models with high densities (n$_{\rm H} = 10^5$ cm$^{-3}$), moderately
warm temperatures (T $= 500$ to 1500 K at the cloud surface), and high
FUV fields (G$_0 = 10^3 - 10^5$ times the mean interstellar field) can
reasonably fit the data.  Figure~\ref{chisq} shows $\chi_\nu^2$
contours for all models projected onto the n$_{\rm H} -$G$_0$ plane.
The best-fit Draine \& Bertoldi model is n2023b, which has n$_{\rm H}$
= 10$^5$ cm$^{-3}$, T = 900 K, and G$_0=5000$.  We fit 22 H$_2$ lines,
excluding 3$-$2 S(2) 2.287 $\mu$m because it appears to be blended
with a strong unidentified nebular line at 2.286 $\mu$m found in
higher-resolution spectra of planetary nebulae (\cite{smith81}).  The
weak high-v transitions are all under-predicted by this model, and
appear to come from lower-density gas (n$_{\rm H} \la 10^3-10^4$
cm$^{-3}$) exposed to a weaker FUV field (G$_0 \la 10^2-10^3$).

The ortho/para ratio of excited H$_2$ determined from the relative
column densities 
%in v=1, J=3 and J=2 
inferred from 1$-$0 S(1) and S(0)
lines is 1.62$\pm$0.07.  This is consistent with the ground state v=0 %,
%J=1 and J=0 
H$_2$ being in LTE with ortho/para ratio of 3 if the FUV
absorption lines populating the non-LTE excited states are optically
thick (\cite{sternberg99}).  Indeed, the best-fit PDR models have
temperatures that are comparable with T$_{\rm rot}$ in the lowest
excited states.

If the extent of the H$_2$ emission indicates that the mean-free path
of a FUV photon is $\sim 200$ pc, then $\langle$n$_{\rm H}\rangle$ = 3
cm$^{-3}$ for a Galactic gas-to-dust ratio, while in the PDR(s)
n$_{\rm H}$ = 10$^4-10^6$ cm$^{-3}$.  This implies that the molecular
gas is extremely clumpy, which is consistent with the range of
densities inferred from the detection of anomalously strong v = 8$-$6
H$_2$ emission.

\section{Conclusions}

The highest surface brightness mid-IR peak in the ISO map of the
Antennae Galaxies is a massive ($\sim 16\times$10$^6$ M$_{\odot}$),
obscured (A$_{\rm V} \sim 9-10$), young (age $\sim 4$ Myr) star
cluster with half-light radius $\sim$ 32 pc, whose strong FUV flux
excites the surrounding molecular ISM on scales of up to 200 pc.  The
cluster spectrum is dominated by extended fluorescently excited
H$_{2}$ emission from clumpy PDRs and nebular emission from compact
\ion{H}{ii} regions.  In contrast, the nearby nucleus of NGC~4039 has a
strong stellar spectrum dominated by cool stars, where the only
emission lines are due to shock-excited H$_2$.  These observations
confirm the potential of near-infrared spectroscopy for exploration
and discovery with the new generation of large ground-based
telescopes.  Our ongoing program of NIRSPEC observations promises
to reveal a wealth of information on the nature of star formation in
star clusters.

\begin{acknowledgements}

We acknowledge the hard work of past and present members of the UCLA
NIRSPEC team: M. Angliongto, O. Bendiksen, G. Brims, L. Buchholz,
J. Canfield, K.  Chin, J. Hare, F. Lacayanga, S. Larson, T. Liu, N.
Magnone, G. Skulason, M. Spencer, J. Weiss and W.  Wong.  We thank
Keck Director Chaffee and all the CARA staff involved in the
commissioning and integration of NIRSPEC, particularly instrument
specialist T. Bida. We especially thank Observing Assistants
J. Aycock, G. Puniwai, C. Sorenson, R. Quick and W. Wack for their
support.  We also thank A. Sternberg for valuable discussions.  We are
grateful to R. Benjamin for providing us with He~{\sc i} emissivity
data.  AMG acknowledges support from a NASA GSRP grant.

\end{acknowledgements}

\end{document}